\documentclass{ws-procs975x65}%
\usepackage{amsmath}
\usepackage{amsfonts}
\usepackage{amssymb}
\usepackage{graphicx}%
\begin{document}

\title{Finite Zero Point Gravitational Energy in the context of Modified Dispersion Relations}
\author{REMO GARATTINI}

\address{Universit\`{a} degli Studi di Bergamo, Facolt\`{a} di Ingegneria,\\ Viale
Marconi 5, 24044 Dalmine (Bergamo) ITALY.\\INFN - sezione di Milano, Via Celoria 16, Milan, Italy\\
\email{remo.garattini@unibg.it}}

\author{GIANLUCA MANDANICI}

\address{Universit\`{a} degli Studi di Bergamo, Facolt\`{a} di Ingegneria,\\ Viale
Marconi 5, 24044 Dalmine (Bergamo) ITALY.\\INFN - sezione di Milano, Via Celoria 16, Milan, Italy\\
\email{gianluca.mandanici@unibg.it}}

\begin{abstract}
We compute the Zero Point Energy in a spherically symmetric background
distorted at high energy as predicted by Gravity's Rainbow. In this context we
setup a Sturm-Liouville problem with the cosmological constant considered as
the associated eigenvalue. The eigenvalue equation is a reformulation of the
Wheeler-DeWitt equation. We find that the ordinary divergences can here be
handled by an appropriate choice of the rainbow's functions, in contrast to
what happens in other conventional approaches.

\end{abstract}

\bigskip\bodymatter

It is quite reasonable to think that General Relativity may have a
modification and a breakdown at the Planck scale. String theory offers a
possible description of the world at that scale but, on the other hand,
introduces a multitude of unsolved problems. One possible approach to see how
General Relativity is distorted at the Planck energy is Gravity's
Rainbow\cite{MS}. Gravity's Rainbow is based on the following assumption:
there exist two unknown functions $g_{1}\left(  E/E_{P}\right)  $ and
$g_{2}\left(  E/E_{P}\right)  $ which have the following property%
\begin{equation}
\lim_{E/E_{P}\rightarrow0}g_{1}\left(  E/E_{P}\right)  =1\qquad\text{and}%
\qquad\lim_{E/E_{P}\rightarrow0}g_{2}\left(  E/E_{P}\right)  =1.\label{lim}%
\end{equation}
They come into play into the usual line element $ds^{2}=g_{\mu\nu}dx^{\mu
}dx^{\nu}$ so that the metric is distorted at the Planck energy $E_{P}$. For
example the spherically symmetric background is distorted in the following way%
\begin{equation}
ds^{2}=-N^{2}\left(  r\right)  \frac{dt^{2}}{g_{1}^{2}\left(  E/E_{P}\right)
}+\frac{dr^{2}}{\left(  1-\frac{b\left(  r\right)  }{r}\right)  g_{2}%
^{2}\left(  E/E_{P}\right)  }+\frac{r^{2}}{g_{2}^{2}\left(  E/E_{P}\right)
}\left(  d\theta^{2}+\sin^{2}\theta d\phi^{2}\right)  \,.
\end{equation}
In a similar manner, the Friedmann-Robertson-Lema\^{\i}tre-Walker metric is
modified by writing%
\begin{equation}
ds^{2}=-\frac{N^{2}\left(  t\right)  }{g_{1}^{2}\left(  E/E_{P}\right)
}dt^{2}+\frac{a^{2}\left(  t\right)  }{g_{2}^{2}\left(  E/E_{P}\right)
}\gamma_{ij}dx^{i}dx^{j}\,,
\end{equation}
where $\gamma_{ij}$ is the metric on the three sphere and $a\left(  t\right)
$ is the scale factor. For small energy compared to $E_{P}$, we recover the
standard General Relativity, but for energy comparable with $E_{P}$, we enter
into the realm of Quantum Gravity. General Relativity provides a natural
quantization scheme based on the Arnowitt-Deser-Misner ($\mathcal{ADM}$)
variables\cite{ADM}, namely the Wheeler-DeWitt (WDW) equation\cite{DeWitt}. A
formal manipulation of the WDW equation with a cosmological term has been
proposed in Ref.\cite{Remo}. The idea is the replacement of the WDW equation
with an eigenvalue equation with $\Lambda/\left(  8\pi G\right)  $ considered
as an eigenvalue computed by means of an expectation value%
\begin{equation}
\frac{1}{V}\frac{\int\mathcal{D}\left[  g_{ij}\right]  \Psi^{\ast}\left[
g_{ij}\right]  \int_{\Sigma}d^{3}x\mathcal{H}\Psi\left[  g_{ij}\right]  }%
{\int\mathcal{D}\left[  g_{ij}\right]  \Psi^{\ast}\left[  g_{ij}\right]
\Psi\left[  g_{ij}\right]  }=\frac{1}{V}\frac{\left\langle \Psi\left\vert
\int_{\Sigma}d^{3}x\hat{\Lambda}_{\Sigma}\right\vert \Psi\right\rangle
}{\left\langle \Psi|\Psi\right\rangle }=-\frac{\Lambda}{8\pi G}.\label{expect}%
\end{equation}
$\mathcal{H}$ is the hamiltonian constraint without the cosmological term.
$\Sigma$ is the hypersurface where we have integrated over and the use of a
functional integration over quantum fluctuation has been performed with the
help of some trial wave functionals of the Gaussian type. $V$ is the volume of
$\Sigma$. The related boundary conditions are dictated by the choice of the
trial wave functionals. The introduction of Gravity's Rainbow avoids a
regularization/renormalization scheme like the one used in Ref.\cite{Remo}.
Its effect on the fundamental pieces composing the WDW equation, namely the
canonical momentum $\pi^{ij}$, the three scalar curvature $R$ and the
super-metric $G_{ijkl}$ is described by\cite{RemoMDR}%
\begin{equation}
\left\{
\begin{array}
[c]{c}%
K_{ij}\rightarrow g_{1}\left(  E/E_{P}\right)  \tilde{K}_{ij}/g_{2}^{2}\left(
E/E_{P}\right)  \\
K\rightarrow g_{1}\left(  E/E_{P}\right)  \tilde{K}\\
\pi^{ij}\rightarrow g_{1}\left(  E/E_{P}\right)  \tilde{\pi}^{ij}/g_{2}\left(
E/E_{P}\right)  \\
\pi\rightarrow g_{1}\left(  E/E_{P}\right)  \tilde{\pi}/g_{2}^{3}\left(
E/E_{P}\right)  \\
R\rightarrow g_{2}^{2}\left(  E/E_{P}\right)  \tilde{R}\\
G_{ijkl}\rightarrow\tilde{G}_{ijkl}/g_{2}\left(  E/E_{P}\right)
\end{array}
\right.  ,\label{rules}%
\end{equation}
where the symbol \textquotedblleft$\sim$\textquotedblright\ indicates the
quantity computed in absence of rainbow's functions $g_{1}\left(
E/E_{P}\right)  $ and $g_{2}\left(  E/E_{P}\right)  $. Then the distorted WDW
equation becomes%
\begin{equation}
\frac{g_{2}^{3}\left(  E/E_{P}\right)  }{\tilde{V}}\frac{\left\langle
\Psi\left\vert \int_{\Sigma}d^{3}x\tilde{\Lambda}_{\Sigma}\right\vert
\Psi\right\rangle }{\left\langle \Psi|\Psi\right\rangle }=-\frac{\Lambda
}{\kappa},\label{WDW3}%
\end{equation}
where%
\begin{equation}
\tilde{\Lambda}_{\Sigma}=\left(  2\kappa\right)  \frac{g_{1}^{2}\left(
E/E_{P}\right)  }{g_{2}^{3}\left(  E/E_{P}\right)  }\tilde{G}_{ijkl}\tilde
{\pi}^{ij}\tilde{\pi}^{kl}\mathcal{-}\frac{\sqrt{\tilde{g}}\tilde{R}}{\left(
2\kappa\right)  g_{2}\left(  E/E_{P}\right)  }\!{}\!.\label{LambdaR}%
\end{equation}
When $E/E_{P}\rightarrow0$, $\tilde{\Lambda}_{\Sigma}\rightarrow\hat{\Lambda
}_{\Sigma}$. With the choice%
\begin{equation}
g_{1}\left(  E/E_{P}\right)  =\left(  1+\beta\frac{E}{E_{P}}\right)
\exp(-\alpha\frac{E^{2}}{E_{P}^{2}}),\qquad g_{2}\left(  E/E_{P}\right)
=1;\qquad\alpha>0,\beta\in\mathbb{R},\label{g1g2}%
\end{equation}
one obtains an interesting result for a de Sitter metric written in a static
form. Indeed, the evaluation of Eq.$\left(  \ref{WDW3}\right)  $ for such a
background leads to the following behavior of $\Lambda/\left(  8\pi G\right)
$%
\begin{equation}
\frac{\Lambda}{8\pi G}\simeq\left\{
\begin{array}
[c]{cc}%
-{\frac{4{\alpha}^{5/2}+3\sqrt{\pi}\beta{\alpha}^{2}}{4\pi^{2}{\alpha}^{9/2}}%
}E_{P}^{4} & x\rightarrow0\\
& \\
E_{P}^{4}\exp(-\alpha x^{2}) & x\rightarrow\infty
\end{array}
\right.  .
\end{equation}
By imposing that%
\begin{equation}
\beta=-{\frac{4\,}{3}}\sqrt{\frac{\alpha}{\pi}},\label{sm}%
\end{equation}
$\Lambda/\left(  8\pi G\right)  $ vanishes for small $x$ and therefore the
result is regular for every value of $x$, where $x=\sqrt{m_{0}^{2}\left(
r\right)  /E_{P}^{2}}$ and where $m_{0}^{2}\left(  r\right)  $ play the
r\^{o}le of an effective mass with%
\begin{equation}
m_{0}^{2}\left(  r\right)  =\frac{6}{r^{2}}-\Lambda_{dS}.\label{potentials}%
\end{equation}
The use of a \textquotedblleft Gaussian\textquotedblright\ form is dictated by
the possibility of doing a comparison with Noncommutative Geometry
models\cite{RG PN}.

\end{document}